%
%
     
\magnification=\magstep1
     
\parskip=3pt plus 1pt minus 1pt
\hfuzz=2pt
\vfuzz=2pt
\tolerance=500
     
\widowpenalty=6000
\displaywidowpenalty=4000
\clubpenalty=6000
     
\hyphenation{Chan-d-ra-sek-har Schwarz-schild an-is-o-tro-pic Max-well
An-to-nov Max-well-ian Hertz-sprung Hertz-sprung-Rus-sell}
\overfullrule=0pt

\catcode`@=11
     
\font\caps=cmcsc10                         
\font\deffont=cmssbx10                     
\font\topicfont=cmbx10                     
\font\subsecfont=cmbx10                    
\font\sectionfont=cmbx10 scaled 1200       
\font\triang=cmmi10 scaled \magstep1       

 \font\syvec=cmbsy10
     
\font\tfont=cmbx10                         
\font\ttfont=cmbx7                         
\font\tttfont=cmbx5                        

  \font\gkvec=cmmib10                          

\font\ninerm=cmr9   \font\ninesy=cmsy9
     
  \font\eightit=cmti8 \font\sixit=cmti7 scaled 857
   \font\fiveit=cmti7 scaled 714
  \font\eightrm=cmr8  \font\eighti=cmmi8
  \font\eightsy=cmsy8 \font\eightbf=cmbx8
     
\font\sixrm=cmr6  \font\fiverm=cmr5
\font\sixi=cmmi6  \font\fivei=cmmi5
\font\sixsy=cmsy6 \font\fivesy=cmsy5
\font\sixbf=cmbx6 
\font\eightex=cmex10 scaled 800
\font\eightdef=cmssbx10 scaled 800

 \font\eightgk=cmmib10 scaled 800
 
\font\eightbsy=cmbsy10 scaled 800

\def\eightpoint{\def\rm{\fam0\eightrm}%
  \textfont0=\eightrm \scriptfont0=\sixrm \scriptscriptfont0=\fiverm%
  \textfont1=\eighti  \scriptfont1=\sixi  \scriptscriptfont1=\fivei%
  \textfont2=\eightsy \scriptfont2=\sixsy \scriptscriptfont2=\fivesy%
  \textfont3=\eightex \scriptfont3=\eightex \scriptscriptfont3=\eightex%
  \textfont\itfam=\eightit \scriptfont\itfam\sixit%
   \scriptscriptfont\itfam\fiveit \def\it{\fam\itfam\eightit}%
  \textfont\bffam=\eightbf \scriptfont\bffam=\sixbf%
    \scriptscriptfont\bffam=\fivebf \def\bf{\fam\bffam\eightbf}%
  \textfont\vecfam=\ttfont \scriptfont\vecfam=\tttfont%
    \scriptscriptfont\vecfam=\tttfont \def\b{\fam\vecfam\ttfont}%
  \def\deffont{\eightdef}%
  \def\topicfont{\eightbf}%
  \def\triangleleft{\hbox{{\teni\char47}}}%
  \def\bnabla{\hbox{{\eightbsy\char114}}}%
  \def\btheta{\hbox{{\eightgk\char18}}}%
  \def\bomega{\hbox{{\eightgk\char33}}}%
  \def\bxi{\hbox{{\eightgk\char24}}}%
  \def\bsigma{\hbox{{\eightgk\char27}}}%
  \normalbaselineskip=9pt
  \normallineskiplimit=0pt
  \normallineskip=0pt
  \parskip=2pt plus 1pt minus 1pt 
  \setbox\strutbox=\hbox{\vrule height7pt depth2pt width0pt}%
  \let\sc=\sixrm  \let\big=\eightbig \let\Big=\eightBig%
  \let\bigg=\eightbigg  \normalbaselines\rm}

\def\eightbig#1{{\hbox{$\textfont0=\eightrm\textfont2=\eightsy
  \left#1\vbox to6.5pt{}\right.\n@space$}}}
\def\eightBig#1{{\hbox{$\textfont0=\eightrm\textfont2=\eightsy
  \left#1\vbox to9.5pt{}\right.\n@space$}}}
\def\eightbigg#1{{\hbox{$\textfont0=\eightrm\textfont2=\eightsy
  \left#1\vbox to11.5pt{}\right.\n@space$}}}

\font\ninerm=cmr9   \font\ninesy=cmsy9 \font\ntfont=cmbx9

  \font\nineit=cmti9 \font\sevenit=cmti7 \font\sixit=cmti7 scaled 857
  \font\ninerm=cmr9  \font\ninei=cmmi9
  \font\ninesy=cmsy9 \font\ninebf=cmbx9
     
\font\sevenrm=cmr7  \font\sixrm=cmr6
\font\seveni=cmmi7  \font\sixi=cmmi6
\font\sevensy=cmsy7 \font\sixsy=cmsy6
\font\sevenbf=cmbx7 
\font\nineex=cmex10 scaled 900
\font\ninedef=cmssbx10 scaled 900
\font\ninebsy=cmbsy10 scaled 900

 \font\ninegk=cmmib10 scaled 900

\def\ninepoint{\def\rm{\fam0\ninerm}%
  \textfont0=\ninerm \scriptfont0=\sevenrm \scriptscriptfont0=\sixrm%
  \textfont1=\ninei  \scriptfont1=\seveni  \scriptscriptfont1=\sixi%
  \textfont2=\ninesy \scriptfont2=\sevensy \scriptscriptfont2=\sixsy%
  \textfont3=\nineex \scriptfont3=\nineex \scriptscriptfont3=\nineex%
  \textfont\itfam=\nineit \scriptfont\itfam\sevenit%
   \scriptscriptfont\itfam\sixit \def\it{\fam\itfam\nineit}%
  \textfont\bffam=\ninebf \scriptfont\bffam=\sevenbf%
    \scriptscriptfont\bffam=\sixbf \def\bf{\fam\bffam\ninebf}%
  \textfont\vecfam=\ntfont \scriptfont\vecfam=\ttfont%
    \scriptscriptfont\vecfam=\tttfont \def\b{\fam\vecfam\ntfont}%
  \def\deffont{\ninedef}%
  \def\topicfont{\ninebf}%
  \def\triangleleft{\hbox{{\teni\char47}}}%
  \def\bnabla{\hbox{{\ninebsy\char114}}}%
  \def\btheta{\hbox{{\ninegk\char18}}}%
  \def\bomega{\hbox{{\ninegk\char33}}}%
  \def\bxi{\hbox{{\ninegk\char24}}}%
  \def\bsigma{\hbox{{\ninegk\char27}}}%
  \normalbaselineskip=10pt
  \normallineskiplimit=0pt
  \normallineskip=0pt
  \parskip=2pt plus 1pt minus 1pt 
  \setbox\strutbox=\hbox{\vrule height7pt depth2pt width0pt}%
  \let\sc=\sevenrm  \let\big=\ninebig \let\Big=\nineBig%
  \let\bigg=\ninebigg
  \normalbaselines\rm%
}

\def\ninebig#1{{\hbox{$\textfont0=\ninerm\textfont2=\ninesy
  \left#1\vbox to6.5pt{}\right.\n@space$}}}
\def\nineBig#1{{\hbox{$\textfont0=\ninerm\textfont2=\ninesy
  \left#1\vbox to9.5pt{}\right.\n@space$}}}
\def\ninebigg#1{{\hbox{$\textfont0=\ninerm\textfont2=\ninesy
  \left#1\vbox to11.5pt{}\right.\n@space$}}}

\def\triangleleft{\hbox{{\triang\char47}}}%
\def\bnabla{\hbox{{\syvec\char114}}}       
\def\btheta{\hbox{{\gkvec\char18}}}        
\def\bomega{\hbox{{\gkvec\char33}}}        
\def\bxi{\hbox{{\gkvec\char24}}}           
\def\bsigma{\hbox{{\gkvec\char27}}}        

\newfam\vecfam

\textfont\vecfam=\tfont \scriptfont\vecfam=\ttfont
\scriptscriptfont\vecfam=\tttfont
\let\b=\vecfont          
     
     

\def\frac#1#2{{\textstyle{#1\over#2}}}

     
\def\spose#1{\hbox to 0pt{#1\hss}}
\def\s{\ifmmode \widetilde \else \~\fi} 
\def\={\overline}
     
\newcount\notenumber
\notenumber=1
\newcount\eqnumber
\eqnumber=1
\newcount\fignumber
\fignumber=1
\newcount\probnumber
\probnumber=1
     
\def\yyskip{\penalty-100\vskip6pt plus6pt minus4pt}
\def\nobreakyyskip{\vskip6pt plus6pt minus4pt}
\def\numberpara{\yyskip\noindent}
\def\nobreaknumberpara{\nobreakyyskip\noindent}

%
\def\note#1{%
  \let\@sf=\empty \ifhmode\edef\@sf{\spacefactor\the\spacefactor}\/\fi
  $\m@th{}^{\the\notenumber}$%
  \insert\footins\bgroup
    \eightpoint\textindent{$\m@th{}^{\the\notenumber}$}%
    \interlinepenalty\interfootnotelinepenalty
    \splittopskip\ht\strutbox 
    \splitmaxdepth\dp\strutbox \floatingpenalty\@MM
    \leftskip\z@skip \rightskip\z@skip \spaceskip\z@skip \xspaceskip\z@skip
    \footstrut #1\strut\par
  \egroup
  \global\advance\notenumber by 1
  \@sf\relax}

\def\f@llrule{\kern-3pt\hrule width \hsize \kern 2.6pt}
\def\p@rtialrule{\kern-3pt\hrule width 2in \kern 2.6pt}
\def\n@rule{}
\let\firstrule=\n@rule
\let\secondrule=\p@rtialrule
\def\footnoterule{\firstrule}
\def\fullfootrules{\let\firstrule=\f@llrule \let\secondrule=\f@llrule%
  \let\footnoterule=\f@llrule}
\def\nofootrules{\let\firstrule=\n@rule \let\secondrule=\p@rtialrule%
  \let\footnoterule=\n@rule}

\def\new{\hbox{\chaphead\the\eqnumber}\global\advance\eqnumber by 1}
\def\ref#1{\advance\eqnumber by -#1 \chaphead\the\eqnumber
     \advance\eqnumber by #1 }
\def\last{\advance\eqnumber by -1 \hbox{\chaphead\the\eqnumber}\advance
     \eqnumber by 1}
\def\eq#1{\advance\eqnumber by -#1 equation \chaphead\the\eqnumber
     \advance\eqnumber by #1}
\def\eqnam#1{\xdef#1{\chaphead\the\eqnumber}}
     
\def\nfig{\chaphead\the\fignumber\ \global\advance\fignumber by 1}
\def\nfiga#1{\chaphead\the\fignumber{#1}\global\advance\fignumber by 1}
\def\rfig#1{\advance\fignumber by -#1 \chaphead\the\fignumber\
     \advance\fignumber by #1}
\def\rfiga#1{\advance\fignumber by -#1 \chaphead\the\fignumber
     \advance\fignumber by #1}
\def\fignam#1{\xdef#1{\chaphead\the\fignumber}}
\def\nprob{\numberpara{\bf\probhead\the\probnumber}.\ \
     \global\advance\probnumber by 1}
\def\nobreaknprob{\nobreaknumberpara{\bf\probhead\the\probnumber}.\ \
     \global\advance\probnumber by 1}
\def\rprob#1{\advance\probnumber by -#1 \probhead\the\probnumber\
     \advance\probnumber by #1}
     
%
\def\sectionbegin#1{\bigbreak\vskip18pt plus 3pt minus 3pt
     {\noindent\interlinepenalty=10000\sectionfont#1\par}
     \nobreak\vskip3pt plus1pt minus1pt
     \relax}
\def\subsectionbegin#1{\bigbreak\vskip18pt plus 3pt minus 3pt
     {\noindent\interlinepenalty=10000\subsecfont#1\par}
     \nobreak\vskip3pt plus 1pt minus 1pt
     \relax}

%
%
     
\newdimen\captiongap  \captiongap=1.0pc  
\newif\ifoutlinefigure  \outlinefigurefalse     

\newif\ifraggedcaptions \raggedcaptionstrue 
     
\def\leftcaption #1{\let\captionjustification=L\let\next=#1\docaption}
\def\rightcaption #1{\let\captionjustification=R\let\next=#1\docaption}
\def\bottomcaption #1{\let\captionjustification=B\let\next=#1\docaption}
     
\def\docaption #1#2#3#4{
  \next 
  \hbox to\z@{\vbox to#2{}\vbox{
    \if B\captionjustification
      \dimen0=\hsize \advance\dimen0 by-#1\relax
      \leftskip=.5\dimen0 \rightskip=\leftskip
      \vskip #2%
      \ifoutlinefigure
        \vbox to\z@{\vss
          \hbox{\hskip\leftskip\vrule\kern-.4pt
            \vbox to#2{\hrule\hbox to#1{\hss}\vss\hrule}\kern-.4pt\vrule}}%
        \nointerlineskip
      \fi
      \vskip\captiongap
    \else\if L\captionjustification
      \leftskip=0pt \rightskip=#1\advance\rightskip by\captiongap
    \else\if R\captionjustification
      \rightskip=0pt \leftskip=#1\advance\leftskip by\captiongap
    \fi\fi\fi
    \ninepoint
    \ifraggedcaptions
      \advance\rightskip by 0pt plus 2em\relax \hyphenpenalty=10000
    \fi
    \noindent{\bf Figure #3.} #4\par
    \if B\captionjustification\else \vskip\captiongap \fi}
    \hss
    \ifoutlinefigure
      \if L\captionjustification
        \hskip\hsize
        \hbox to\z@{\hss\vrule\kern-.4pt
            \vbox to#2{\hrule\hbox to#1{\hss}\vss\hrule}\kern-.4pt\vrule}%
        \hskip-\hsize
      \fi
      \if R\captionjustification
        \hbox to\z@{\vrule\kern-.4pt
            \vbox to#2{\hrule\hbox to#1{\hss}\vss\hrule}\kern-.4pt\vrule\hss}%
      \fi
    \fi
    }
    \vskip 2pt
  \endinsert 
}
     
\def\lta{\mathrel{\spose{\lower 3pt\hbox{$\mathchar"218$}}
     \raise 2.0pt\hbox{$\mathchar"13C$}}}
\def\gta{\mathrel{\spose{\lower 3pt\hbox{$\mathchar"218$}}
     \raise 2.0pt\hbox{$\mathchar"13E$}}}

     
\newif\ift@p \newif\ifp@ge \newif\if@mid
\newinsert\botins
\skip\botins=\medskipamount \dimen\botins=\maxdimen \count\botins=1000
\newinsert\topins
\skip\topins=\medskipamount \dimen\topins=\maxdimen \count\topins=1000
\newskip\interinsertskipamount \interinsertskipamount=\medskipamount
     
\def\botinsert {\t@pfalse\@midfalse\p@gefalse\@ins}
\def\topinsert {\t@ptrue \@midfalse\p@gefalse\@ins}
\def\midinsert {\t@ptrue \@midtrue \p@gefalse\@ins}
\def\pageinsert{\t@ptrue \@midfalse\p@getrue \@ins}
     
\def\@ins{\begingroup\setbox\z@\vbox\bgroup} 
\def\endinsert{\egroup 
  \if@mid
    \dimen@=\ht\z@ \advance\dimen@ by\dp\z@
    \advance\dimen@\bigskipamount \par\advance\dimen@\pagetotal
    \ifdim\dimen@>\pagegoal\@midfalse\p@gefalse\fi
  \fi
  \if@mid
    \bigskip\box\z@\bigbreak
  \else
    \ift@p \insert\topins
    \else  \insert\botins \fi
    {\penalty100 
    \splittopskip\z@skip \splitmaxdepth\maxdimen \floatingpenalty\z@
    \ifp@ge \dimen@\dp\z@ \vbox to\vsize{\unvbox\z@\kern-\dimen@}
    \else \box\z@\nobreak\vskip\interinsertskipamount\fi}%
  \fi\endgroup}
     
\def\pagecontents{%
  \ifvoid\topins\else \unvbox\topins\vskip\skip\topins\fi
  \dimen@=\dp\@cclv \unvbox\@cclv 
  \ifvoid\botins\else \vskip\skip\botins\unvbox\botins\fi 
  \ifvoid\footins\else 
    \vskip\skip\footins
    \footnoterule
    \unvbox\footins\fi
  \ifr@ggedbottom \kern-\dimen@ \vfil \fi
  \ifnum\insertpenalties>0\xdef\footnoterule{\secondrule}\else%
  \xdef\footnoterule{\firstrule}\fi
}
     
\catcode`\@=12
     
\def\chaphead{}

\newbox\abstr

\def\numberpara{\yyskip\noindent}

\def\foot#1{\raise3pt\hbox{\eightrm \the\notenumber}
     \hfil\par\vskip3pt\hrule\vskip6pt
     \noindent\raise3pt\hbox{\eightrm \the\notenumber}
     #1\par\vskip6pt\hrule\vskip3pt\noindent\global\advance\notenumber by 1}

\def\abstract#1{\setbox\abstr=\vbox{\hsize 5.0truein{\par\noindent#1}}
    \centerline{ABSTRACT} \vskip12pt \hbox to \hsize{\hfill\box\abstr\hfill}}



\vsize 9truein \hsize 6.71truein
\voffset=-0.25truein \hoffset=-0.21truein

\

\centerline{\rm ESTABLISHING THE CONNECTION BETWEEN PEANUT-SHAPED BULGES}
\centerline{\rm AND GALACTIC BARS}
\bigskip
\centerline{\caps Konrad Kuijken\note{present
address: Kapteyn Instituut, PO Box 800, 9700 AV Groningen, the Netherlands}}
\centerline{\eightpoint Harvard-Smithsonian Center for Astrophysics, 
60 Garden Street, Cambridge, MA 02138. }
\centerline{\eightpoint I: kuijken@cfa.harvard.edu} 
\medskip 
\centerline{\caps and}
\medskip
\centerline{\caps Michael R.\ Merrifield}
\centerline{\eightpoint Department of Physics, 
University of Southampton, Highfield, SO9 5NH, Britain. }
\centerline{\eightpoint I: mm@phastr.soton.ac.uk} 


\bigskip

\sectionbegin{\rm ABSTRACT}

It has been suggested that the peanut-shaped bulges seen in some edge-on disk
galaxies are due to the presence of a central bar.  Although bars cannot be
detected photometrically in edge-on galaxies, we show that barred potentials
produce a strong kinematic signature in the form of double-peaked
line-of-sight velocity distributions with a characteristic ``figure-of-eight''
variation with radius.  We have obtained spectroscopic observations of two
edge-on galaxies with peanut-shaped bulges (NGC~5746 and NGC~5965), and they
reveal exactly such line-of-sight velocity distributions in both their gaseous
(emission line) and their stellar (absorption line) components.  These
observations provide strong observational evidence that peanut-shaped bulges
are a by-product of bar formation.

\medskip
\noindent {\it Subject headings:} 
galaxies: spiral --- galaxies: kinematics and dynamics --- galaxies: structure
--- galaxies: individual (NGC~5746, NGC~5965)

\vfill\eject
\bigskip

\sectionbegin{\rm 1. INTRODUCTION}

Around $30\%$ of disk galaxies are observed to be barred (Sellwood \&
Wilkinson 1993).  Normal mode analysis (e.g. Hunter 1992) and N-body
simulations (e.g.  Sparke \& Sellwood 1987) show that a global instability in
self-gravitating disks provides a natural explanation for the formation of
such structures.  Full three-dimensional numerical simulations also show that
the formation of a bar is accompanied by a buckling instability, which means
that barred systems can take on a peanut shape when viewed from the side
(Combes \& Sanders 1981, Combes et al. 1990, Raha et al.~1991).  By
comparison, at least 20\% of edge-on disk galaxies have boxy or peanut-shaped
bulge isophotes (Shaw 1987).  These percentages are sufficiently similar for
it to be very tempting to conclude that all the peanut-shaped bulges are the
product of central bars.

Unfortunately, the link between these two phenomena has proved diffult to
establish observationally: the boxy isophotes in a galaxy's bulge are only
observable photometrically if the system is close to edge on, but the presence
of a bar is only generally apparent if the galaxy is viewed in a more face-on
orientation.  In the case of the Milky Way, where the bulge has a slight
peanut shape (Weiland et al.~1994), perspective effects provide a convincing
demonstration that the Galaxy also contains a bar (Blitz \& Spergel 1991), but
these effects are far too small to be seen in external galaxies. The only
galaxy with clear photometric evidence for both a boxy bulge and a bar is
NGC~4442 (Bettoni \& Galletta 1994): in this case the boxy distortions are
sufficiently strong that they are still apparent at an inclination of
$72^\circ$. More circumstantial evidence for a link is provided by NGC~1381
(de Carvalho \& da Costa 1987), an edge-on S0 galaxy with a box-shaped bulge
and a central plateau in its radial distribution of light which is suggestive
of a central bar.  Similarly, for a large sample of edge-on galaxies Dettmar
\& Barteldrees (1990) find an association between box- or peanut-shaped bulges
and a thin central component in the light distribution, which they interpret
as possible edge-on bars.  Finally, some nearly edge-on systems with
peanut-shaped bulge isophotes such as NGC~5746 and NGC~5965 appear to have
bulges which are slightly tipped with respect to the disk plane: one possible
interpretation for this effect is a bar seen not quite end-on.  However, none
of these data establish a strong link between boxy-bulge galaxies and barred
ones.

Kinematic data have also failed to provide the definitive answer to date: the
only case of an edge-on, box/peanut-bulge galaxy for which there has been
convincing evidence for non-axisymmetric central dynamics is, once again, the
Milky Way, where the CO and HI kinematics show the characteristics of a bar
(Binney et al.~1991). In external galaxies, it has been shown that
peanut-shaped bulges tend to rotate cylindrically (i.e. with azimuthal
velocities which are constant with height; Jarvis 1987, 1990) whereas the
rotational velocity of round bulges declines with distance from their
galaxies' planes (e.g. Kormendy \& Illingworth 1982). The cylindrical rotation
is consistent with the predictions of N-body simulations of bars, but the
evidence is still circumstantial: there also exist simple two-integral
axisymmetric models for boxy bulges which exhibit cylindrical rotation much
like the observations (Rowley 1988).

In this {\it Letter} we illustrate how the presence of a central bar produces
characteristic structure in the line-of-sight velocity distribution (LOSVD) of
an edge-on galaxy (\S 2).  This signature would not occur in an axisymmetric
potential, and so it provides unequivocal evidence for the existence of a bar.
In \S 3, we present spectra of two galaxies with peanut-shaped bulges which
show exactly the predicted LOSVDs.  We can therefore conclude that these
galaxies do, indeed, contain hidden bars.

\sectionbegin{\rm 2. THE KINEMATIC SIGNATURE OF A BAR}

The kinematics of a galaxy can be understood qualitatively from study of the
closed orbits allowed by the potential. The collisional nature of interstellar
gas means that these orbits are a good approximation to the motions of this
component.  Stellar motions are somewhat more complicated, since stars have
the additional degree of freedom that they can follow orbits which oscillate
about the closed orbits, and there may also be significant populations of
stars on irregular, chaotic orbits (e.g., Sellwood \& Wilkinson 1993).
Nevertheless, in a potential with a reasonable density of non-chaotic orbits,
we would expect features which show up in the distribution of closed orbits to
also appear in the stellar motions, perhaps somewhat broadened by the stars'
oscillations about these orbits and diluted by their superposition on any
irregular component.

Figure~1({\it a}) shows the observable kinematic properties --- LOSVD as a
function of projected radius --- for the closed orbits out to a finite radius
in an axisymmetric potential.  As long as the density of material is falling
with radius, then projection along the line of sight at a radius $R$ produces
a high observed density of material moving at the local circular speed, $
v_{los} = v_c(R)$, together with a tail of material moving with lower
line-of-sight velocities.

The orbits in a barred galaxy have been discussed extensively in the
literature: examples include the articles by Contopoulos \& Papayanopoulos
(1980) and Athanassoula (1993). Inside the inner Lindblad resonance (ILR) in a
barred potential, the closed orbits are aligned perpendicular to the
bar. Between the ILR and the corotation radius (the radius at which stars
orbit the galaxy at the same speed at which the bar's pattern rotates), the
orbits are distorted with their major axes aligned along the bar. In a broad
region around corotation there are no closed, non-self-intersecting orbits
available, and at larger radii still the orbits are only slightly elongated
and aligned perpendicular to the bar. Because gas orbits cannot intersect each
other or themselves, the gas is naturally forced onto the closed
non-overlapping orbits allowed by the potential. Figure~1({\it b}) shows these
orbits projected onto the $\{R, v_{los}\}$ plane for a particular bar model,
viewed at an angle intermediate between the bar axes. Unlike the LOSVD
displayed in Figure 1({\it a}), this diagram is rich in features. The
transitions between the different orbit families result in gaps in the
diagram; the prominent gaps which give the distribution its
``figure-of-eight'' appearance reflect the lack of available orbits near the
corotation radius.

\sectionbegin{\rm 3. OBSERVATIONS OF NGC~5746 AND NGC~5965}

I-band CCD images of two edge-on galaxies with peanut-shaped bulges, NGC~5746
and NGC~5965, were obtained using the FLWO 48-inch telescope.  These images,
presented in Figs~2({\it a}) and 3({\it a}), clearly show the distorted
isophotes of the bulges in these systems.

In order to look for the kinematic signature of the bars which may lie hidden
in these two galaxies, we obtained long slit spectra along their major axes.
The data were obtained using the ISIS two-armed spectrograph on the 4.2m
William Herschel Telescope on the nights of 1993 June 11--12 through a slit of
width 1.25 arcseconds. Tektronix CCD detectors and 1200 line/mm gratings were
used in both arms of the spectrograph, producing spectra at a resolution of
0.41\AA/pixel, with 2.3 pixels FWHM. In order to reduce readout noise, the
data were binned on the CCDs in the spatial direction to produce a spatial
resolution of 1.05~arcseconds/pixel. The red arm of the spectrograph was used
to obtain spectra centered on either the H$\alpha$ line (6563\AA) or the
triplet of CaII lines at $\sim 8600$\AA. Spectra from the blue arm were
centered on the Mg~b feature (5190\AA).  Total exposure times on these objects
were 3600s on NGC~5746 and 4800s on NGC~5965.

The spectra from both arms were initially reduced using standard IRAF longslit
procedures: bias subtraction, flatfielding, two-dimensional re-binning to a
logarithmic wavelength scale based on calibration from bracketing Ne-Ar lamp
exposures, and co-addition of multiple observations. Significant fringing in
the spectra of the calcium triplet was also flatfielded out using continuum
lamp exposures obtained during the course of the night. Both arms of the ISIS
spectrograph have remarkably linear spectral dispersions, and low order
polynomial fits produced typical RMS residuals in the wavelength solution of
$\sim 0.015$\AA.

\subsectionbegin{3.1 \it Emission Line Spectra}

The spectra of the [NII] line at 6583\AA\ from the two galaxies are presented
in Figs. 2({\it b}) and 3({\it b}).  We show this line from the spectral
region around H$\alpha$ because it does not lie on top of any stellar
absorption features (such as H$\alpha$ itself), and so the stellar continuum
has been effectively removed from each spectrum by simply subtracting the mean
galaxy luminosity profile.  It is apparent that the emission lines are split
into two components forming the distorted figure-of-eight shape predicted by
the barred galaxy model [c.f. Fig.~1({\it b})]. Axisymmetric gas orbits, which
project as straight lines in this plane [Fig.~1({\it a})] could not combine to
produce these plots.

Figures~2({\it b}) and 3({\it b}) also show the circular rotation speeds of
these galaxies obtained from the CCD images using Merrifield's (1991)
quadrature which assumes that the galaxies are axisymmetric.  This method also
assumes that the galaxies have a constant mass-to-light ratio, which has been
shown to be a valid assumption in the central parts of other disk galaxies
(e.g. Kent 1986).   In each case, the value of the mass-to-light ratio was
chosen to match the observed rotational velocity at large radii.  At small
radii, neither emission line component matches the shape of this rotation
curve (as they would in an axisymmetric galaxy), but they bracket the curve as
predicted by the bar model.

The photometry in Figs.~2({\it a}) and 3({\it a}) also shows that the
peanut-shaped distortions in NGC~5746 and NGC~5965 extend to $\sim 35$ and
$\sim 15$ arcseconds respectively.  If these distortions are due to bars, then
we can associate these distances with the ends of the bars and hence,
presumably, the bars' corotation radii.  These numbers fit nicely with
the LOSVDs presented in Figs.~2({\it b}) and 3({\it b}), as the split LOSVDs
extend to approximately twice the corotation radius in a barred potential
[c.f.~Fig.~1({\it b})].

\subsectionbegin{3.2 \it Absorption Line Spectra}

The reduced absorption line spectra were analyzed using the unresolved
gaussian decomposition (UGD) algorithm described by Kuijken \& Merrifield
(1993).  This method extracts the stellar LOSVD from the Doppler broadening
and shift in spectral lines by modeling the velocity distribution as the sum
of a set of unresolved gaussian distributions with fixed means and
dispersions. The best fit LOSVD is then found by convolving this model with
spectra of standard stars (obtained on the same observing run with the same
spectral set-up) and varying the amplitudes of the gaussian components until
the galaxy spectrum is best reproduced in a least-squares sense. For the
current analysis, we coadded data along the longslit spectrum until we
obtained a signal-to-noise ratio of at least 20 per wavelength pixel in each
spatial bin. Each LOSVD was then modeled as the sum of 21 gaussian components
with means separated by $75\ {\rm km}\ {\rm s}^{-1}$, and dispersions of $50\
{\rm km}\ {\rm s}^{-1}$.

Figures 2({\it c,d}) and 3({\it c,d}) show the stellar LOSVDs derived from
this analysis.  Deconvolution of a galaxy spectrum to extract the LOSVD is an
intrinsically noise-amplifying process.  We therefore expect the derived
LOSVDs to be somewhat noisy, even with the high signal-to-noise ratio spectra
obtained from these observations.  A good measure of the reliability of the
results can be obtained by comparing the LOSVDs obtained from the Mg~b spectra
taken with the blue arm of the spectrograph with those obtained from
absorption line spectra taken with its red arm. In the case of NGC~5746, these
red arm spectra were centered on the calcium triplet; for NGC~5965, the
spectra used were all centered on the H$\alpha$ line (with the emission lines
automatically rejected from the UGD fitting process [Kuijken \& Merrifield
1993]). The two arms of the spectrograph are essentially independent
instruments which obtain spectra with different CCD responses, different
velocity resolutions, different contamination from night sky lines, etc.
There are therefore not likely to be any strong common systematic problems,
and so differences between the derived LOSVDs can be attributed to the
uncertainty due to the finite signal-to-noise ratio of the data.  By comparing
parts ({\it c}) and ({\it d}) of Figs.~2 and 3, it is apparent that even quite
subtle structure in the LOSVDs is reproduced in these independent data sets.
In particular, the clear split into two components, one faster than the
circular speed and one slower than the circular speed, mirrors the behavior of
the emission line spectra.

The similarity between the LOSVDs derived from different parts of the spectrum
also allows us to address any concerns about the effects of extinction in
these galaxies.  Inspection of the photometry in Figs.~2({\it a}) and 3({\it
a}) reveals that, although these disk galaxies are of early Hubble types, they
do show some obscuration by dust lanes.  The galaxies' inclinations mean that
the lines of sight of our spectral observations will intersect the dust lanes
mid-way through each galaxy.  The observed structure in the LOSVDs might
therefore be the result of selective dust extinction along these lines of
sight through intrinsically axisymmetric galaxies.  However, the close
agreement between the LOSVDs derived from the red and blue spectral data
belies this explanation.  In the case of NGC~5746, for example, the sizes of
the high velocity components (which would come from material at the smallest
radii in an axisymmetric galaxy) are identical to within the observational
errors of $\sim 30\%$.  From the standard interstellar extinction curve
(e.g.~Mihalas \& Binney 1981), the difference between the extinction at
5200\AA\ and that at 8600\AA\ can only be less than $30\%$ if the extinction
at 8600\AA\ is less than 0.4 magnitudes.  No matter how such a modest amount
of extinction is distributed along the line of sight, it is nowhere near
enough to explain the gap between the slow and fast components in the LOSVD,
where the projected phase density is depressed by at least a factor of 5.

\sectionbegin{\rm 4. DISCUSSION}

An edge-on barred galaxy displays a characteristic kinematic signature in that
its LOSVD has two high density peaks.  One of these peaks travels faster than
the local circular speed, the other more slowly, and the variation in their
velocities with projected radius along the major axis forms a distorted
figure-of-eight in the $\{R, v_{los}\}$ plane. This signature allows edge-on
barred galaxies to be identified. We have used this indicator in optical
spectra to show that the two galaxies with peanut-shaped bulges which we have
studied --- NGC~5746 and NGC~5965 --- also harbor bars. This discovery
provides strong observational support for the hypothesis that peanut-shaped
and boxy bulges are a by-product of bar formation.

Binney et al.~(1991) have demonstrated that CO and HI observations reveal the
presence of a bar at the center of the Milky Way.  We might therefore expect
that data in these wavebands from external galaxies should also show the sort
of signature that we have found in optical observations.  Indeed, CO
observations of the edge-on boxy-bulged S0 galaxy NGC~4710 (Wrobel \& Kenney
1992) show exactly the same two-component velocity distribution, and we would
suggest this observation implies that NGC~4710 also contains a bar.

With the current generation of 4-meter telescopes and efficient spectrographs,
it is possible to obtain spectra of the central parts of external galaxies at
high signal-to-noise ratios.  As we have shown, such spectra are of sufficient
quality to study quite fine kinematic structure in both the gaseous and
stellar components of these systems.  Further comparison of kinematic data
from peanut-shaped bulges with detailed dynamical models of bars will enable
us to explore the connection between these two phenomena in much greater
depth.

\bigskip
KK was supported by a Hubble Fellowship through grant HF-1020.01-91A awarded
by the Space Telescope Science Institute (which is operated by the Association
of Universities for Research in Astronomy, Inc., for NASA under contract
NAS5-26555). MM was supported by a PPARC Advanced Fellowship
(B/94/AF/1840). The spectral observations were obtained with the William
Herschel Telescope, operated on the island of La Palma by the Royal Greenwich
Observatory in the Spanish Observatorio del Roque de los Muchachos of the
Instituto de Astrofisica de Canarias. We thank Reynier Peletier for his
assistance with the observations, and Marijn Franx for a thorough reading of
the manuscript.

\sectionbegin{\rm REFERENCES}

\def\refitem{\par\parskip 0pt\noindent\hangindent 20pt}

\refitem{Athanassoula, E., MNRAS, 259, 328}
\refitem{Bettoni, D. \& Galletta, G. 1994, A\&A, 281, 1}
\refitem{Binney, J., Gerhard, O.E., Stark, A.A., Bally, J. \& Uchida,
K.I. 1991, MNRAS, 252, 210}
\refitem{Binney, J. \& Tremaine, S. 1987, Galactic Dynamics
(Princeton: Princeton University Press)}
\refitem{Blitz, L. \& Spergel, D.N. 1991, ApJ, 379, 631}
\refitem{Combes, F., Debbasch, F., Friedli, D. \& Pfenniger, D. 1990, A\&A,
233, 82}
\refitem{Combes, F. \& Sanders, R.H. 1981, A\&A, 96, 164}
\refitem{Contopoulos, G. \& Papayannopoulos, Th.D., 1980, A\&A, 92, 33}
\refitem{de Carvalho, R.R. \& da Costa, L.N. 1987, A\&A, 171, 66}
\refitem{Dettmar, R.-J. \& Barteldrees, A. 1990, in ESO/CTIO workshop on bulges
of galaxies,  259.}
\refitem{Hunter, C. 1992, in Astrophysical Disks, eds S.F. Dermott, J.H. Hunter
\& R.E. Wilson (New York: New York Academy of Sciences), 22}
\refitem{Jarvis, B.J. 1987, AJ, 94, 30}
\refitem{Jarvis, B.J. 1990, in Heidelberg Conference on the Dynamics and
interactions of galaxies, ed. R. Wielen (Springer), 416}
\refitem{Kent, S.M. 1986, AJ, 91, 1301}
\refitem{Kormendy, J. \& Illingworth, G. 1982, ApJ, 256, 460}
\refitem{Kuijken, K. \& Merrifield, M.R. 1993, MNRAS, 264, 712}
\refitem{Merrifield, M.R. 1991, AJ, 102, 1335}
\refitem{Mihalas, D. \& Binney. J. 1981, Galactic Astronomy 
        (New York: Freeman)}
\refitem{Mulder, W.A. \& Liem, B.T. 1986, A\&A, 157, 148}
\refitem{Raha, N., Sellwood, J.A., James, R.A. \& Kahn, F.D. 1991, Nature, 352,
411}
\refitem{Rowley, G. 1988, ApJ, 331, 124 }
\refitem{Sellwood, J.A. \& Wilkinson, A. 1993, Rep Prog Phys, 56, 173}
\refitem{Shaw, M.A. 1987, MNRAS, 229, 691}
\refitem{Sparke, L.S. \& Sellwood, J.A. 1987, MNRAS,  225, 653}
\refitem{Weiland, J.L. et al.~1994, ApJ, 425, L81}
\refitem{Wrobel \& Kenney 1992, ApJ, 399, 94}

\vfill\eject
\sectionbegin{\rm FIGURE CAPTIONS}

{\caps Fig.~1}---Orbits projected into the observable $\{R, v_{los}\}$
plane for an edge-on galaxy with potential
$\Psi(R,\theta)=0.5\ln(0.2^2+r^2)-[1+r^2(1+\epsilon\cos2\theta)]^{-1.5}$
at an intermediate position angle. Orbits are randomly sampled, and
plotted with symbols of sizes which mimic a radial exponential density
law of unit scale length. ({\it a}) axisymmetric potential,
$\epsilon=0$, and ({\it b}) barred potential, $\epsilon=0.2$. The
pattern speed is such that the corotation radius of the barred
potential occurs at a radius of unity.

{\caps Fig.~2}---Photometry and kinematics of NGC~5746. ({\it a}) I-band
photometry of the galaxy, with contours spaced by 0.3 magnitudes: the thick
line indicates the position of the slit for the spectral observations.
Line-of-sight velocity distribution, normalized to unit velocity integral, as
a function of projected radius for: ({\it b}) the gaseous component as traced
by the [NII] line at 6583\AA; ({\it c}) the stellar component as derived from
absorption lines around the Mg~b feature at 5170\AA; and ({\it d}) the stellar
component as derived from absorption lines around the Ca triplet at 8600\AA.
The greyscale bar indicates the relative phase space density scale in both
({\it c}) and ({\it d}).

{\caps Fig.~3}---Photometry and kinematics of NGC~5965. ({\it a}) I-band
photometry of the galaxy, with contours spaced by 0.3 magnitudes: the thick
line indicates the position of the slit for the spectral observations; the
diagonal feature is due to a bad column in the CCD.  Line-of-sight velocity
distribution, normalized to unit velocity integral, as a function of projected
radius for: ({\it b}) the gaseous component as traced by the [NII] line at
6583\AA; ({\it c}) the stellar component as derived from absorption lines
around the Mg~b feature at 5170\AA; and ({\it d}) the stellar component as
derived from absorption lines around the H$\alpha$ line at 6563\AA.  The
greyscale bar indicates the relative phase space density scale in both ({\it
c}) and ({\it d}).  The absence of strong absorption features and
contamination by emission lines limit the signal-to-noise ratio in ({\it d}).

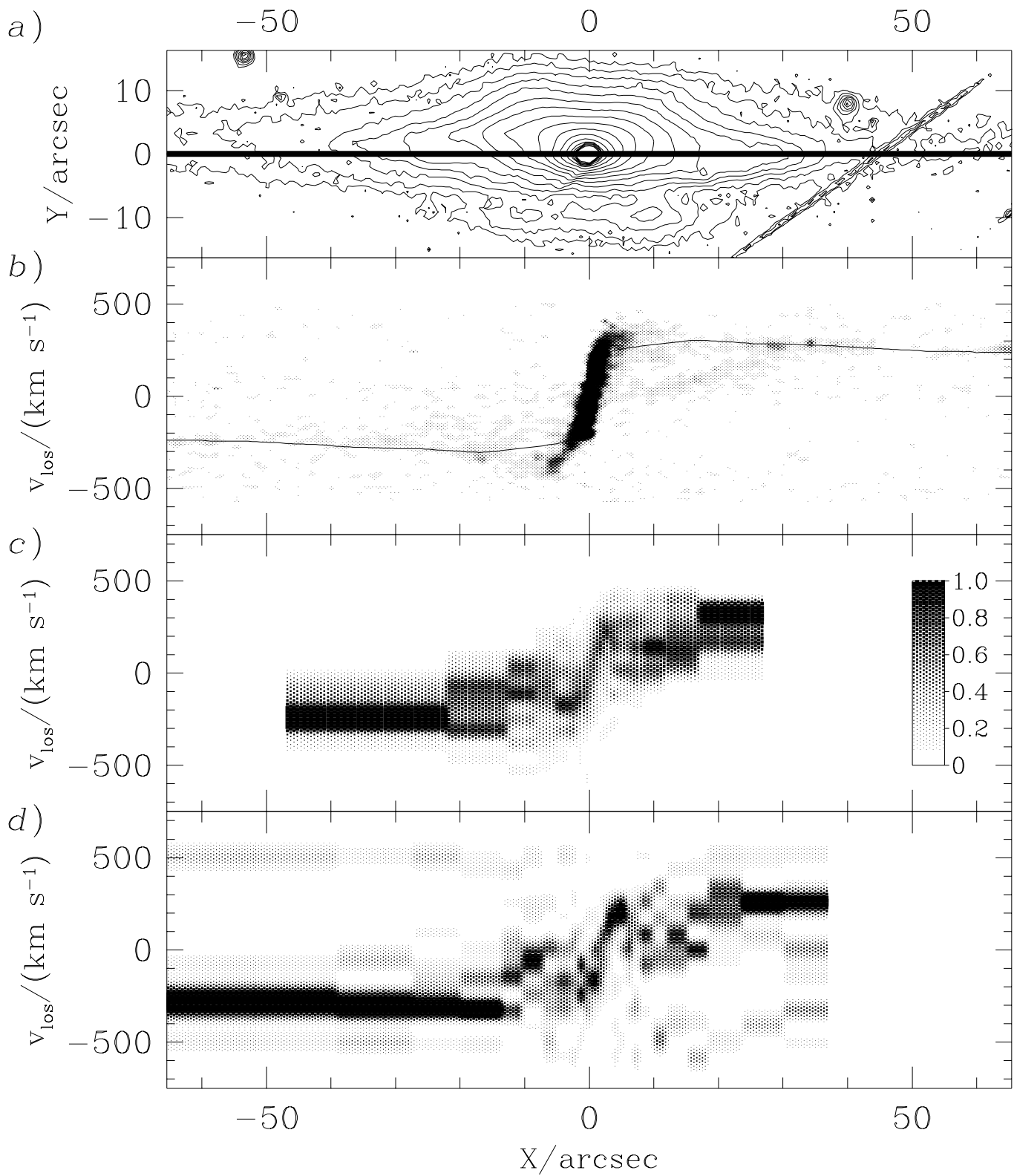
\bye